\documentclass[%
 reprint,
 amsmath,amssymb,
 aps,
 nofootinbib
 prb,
 superscriptaddress
 ]{revtex4-1}
\usepackage{scrextend}
\usepackage{graphicx}
\usepackage{dcolumn}
\usepackage{bm}
\usepackage{tabularx}
\usepackage[utf8]{inputenc}
\usepackage[english]{babel}
\usepackage[bottom]{footmisc}
\usepackage{color}

\usepackage{amsmath}
\usepackage{soul,xcolor}
\setstcolor{red}
\usepackage{hyperref}
\usepackage{booktabs}
\usepackage{epstopdf}

\DeclareMathAlphabet{\pazocal}{OMS}{zplm}{m}{n}

\begin{document}

\preprint{APS/123-QED}

\title{Designing new Zintl phases SrBaX (X = Si, Ge, Sn)
for thermoelectric applications using \textit{ab initio} techniques}
\author{Vivek Gusain}
\author{Mohd Zeeshan}
\author{B. K. Mani}
\email{bkmani@physics.iitd.ac.in}
\affiliation {Department of Physics, Indian Institute of Technology,
Hauz Khas, New Delhi 110016, India}

\date{\today}

\begin{abstract}
Slack's phonon-glass and electron-crystal concept has been
the guiding paradigm for designing new thermoelectric materials.
Zintl phases, in principle, have been shown as great contenders
of the concept and thereby good thermoelectric candidates. With
this as motivation, we design new Zintl phases SrBaX (X = Si, Ge,
Sn) using state-of-the-art computational methods. Herein, we use
first-principles simulations to provide key theoretical insights
to thermal and electrical transport properties. Some of the key
findings of our work feature remarkably low lattice thermal
conductivities ($<$~1~W~m$^{-1}$~K$^{-1}$), putting proposed
materials among the well-known thermoelectric materials such
as SnSe and other contemporary Zintl phases. We ascribe such
low values to antibonding states induced weak bonding in the
lattice and intrinsically weak phonon transport, resulting
in low phonon velocities, short lifetimes, and considerable
anharmonic scattering phase spaces. Besides, our results on
electronic structure and transport properties reveal tremendous
performance of SrBaGe ($ZT\sim$ 2.0 at 700~K), highlighting
the relevance among state-of-the-art materials such as SnSe.
Further, the similar performances for both $p$- and $n$-type
dopings render these materials attractive from device fabrication
perspective. We believe that our study would invite experimental
investigations for realizing the true thermoelectric potential
of SrBaX series.


\end{abstract}

\maketitle

\section{Introduction}

Thermoelectricity is one of the promising technologies which
could shape the modern world \cite{Singh24}. It can change the
way we use our energy resources and could be revolutionary in
future power generation \cite{Garmroudi21}. The working principle
of a thermoelectric device hinges on the conversion of environment's
waste heat into usable electrical energy \cite{Wehbi22}. An efficient
thermoelectric material can power sensors \cite{Xie23}, wearables
\cite{Jia21}, batteries \cite{Wang18}, communication devices \cite{Zoui20},
and remote stations \cite{Belsky23}, depending on its power output.
The efficiency of a thermoelectric material is commonly expressed
in terms of figure of merit, i.e., $ZT = S^2 \sigma T/\kappa$,
where $S$, $\sigma$, and $\kappa$ are Seebeck coefficient, electrical
conductivity, and thermal conductivity which includes both electronic
($\kappa_e$) and thermal ($\kappa_L$) contributions \cite{Karmakar20,
Maassen21, Berry17}. Needless to say, a good thermoelectric material
should possess a high $S$, high $\sigma$, and low $\kappa$. Unfortunately,
these parameters are connected with each other posing a tricky optimization
challenge \cite{Boona17, Lee17, Bell08}.

Slack \cite{Slack95}, in 1995, proposed that a good thermoelectric material
should behave like a crystal which eases the flow of current, and simultaneously
it should have glass like properties, imparting low $\kappa_L$. Such features
demand narrow band gap semiconductors with high mobility of charge carriers
and minimum thermal conductivity. The formidable requirements are most closely
followed by systems comprising  elements with small electronegativity differences
and heavier atomic masses like Bi, Te, Sb, Sn, Se, and Ge. Likewise Bi$_2$Te$_3$
\cite{Serrano17} and SnSe \cite{Zhao14} are among the best thermoelectric
materials with excellent figure of merit values. The small electronegativity
difference renders narrow band gaps, suitable for mobility of charge carriers.
On the other hand, heavier elements vibrate slowly and thereby exhibit low phonon
velocities, adequate for low $\kappa_L$. Slack's concept, also known as PGEC
(\textit{phonon-glass and electron-crystal}), has been the cornerstone in modern
day research in thermoelectrics \cite{Narducci24}.

Though not always necessary, the PGEC requirements make it customary to look
for a ternary material at least. In a ternary material, two atoms should form
covalent like framework which could facilitate the movement of charge carriers,
whereas third atom should have ionic type linkage with the rest framework,
conducive to low $\kappa_L$. There might be an another possibility that two
atoms collectively donate electrons to third, which in turn form isolated
anionic chains. The search for such exquisite features in one material narrows
down the target to Zintl phases which have been demonstrated to closely follow
the PGEC concept to a certain extent \cite{Wang20, Balvanz20, Chen19}. A
classical description of Zintl phases can be understood in terms of a
hypothetical ABX compound, where B and X form a polyanionic framework
using electrons donated by A, i.e., A$^{n+}$[BX]$^{n-}$. Thuswise,
[BX]$^{n-}$ can help in electrons movement through lattice and A$^{n+}$
can scatter heat carrying phonons via different mechanisms such as
resonant (localized) phonon scattering by rattling modes \cite{Lin16},
mass fluctuation \cite{Abeles63}, and strain field scatterings
\cite{Yang04}.

Based on the PGEC guidelines as discussed and available literature,
our aim is to design new thermoelectric materials using state-of-the-art
computational techniques. First-principles simulations give this liberty
to design materials by handpicking the desirable elements, provided the
material should withstand the rigorous test of its stability. We scanned
the periodic table with a zest for unreported materials which could closely
abide by the PGEC concept. Such prerequisites narrowed down our search to
ternary unsynthesized materials comprising heavier earth-abundant elements,
and fulfilling the Zintl phase criteria. After testing various possibilities,
we arrived at SrBaSn, which fits well our requirements. It is not reported,
comprises heavy elements like Ba and Sn
which are likely to lower the acoustic phonon frequencies. The heavy elements
like Ba and Sn when combined with Sr can improve phonon scatterings on account
of mass contrast. Simultaneously, Sr and Ba can also scatter heat carrying
phonons by acting as rattler modes. Further, there seems to be a possibility
of narrow or moderate band gap in SrBaSn, owing to electronegativity differences
between the constituent elements, i.e., $\sim$1 for Sr-Sn, Ba-Sn, and 0.06 for
Ba-Sn. In addition, SrBaSn can be classified as 1-1-1 type Zintl
phases which have gained attention in recent years. With compounds
like SrAgSb \cite{Zhang20} ($ZT\sim0.5$ at 773~K), BaCuSb \cite{Zheng22}
($ZT\sim0.4$ at 1010~K), and NaCdSb \cite{Guo23} ($ZT\sim1.3$ at 673~K),
the 1-1-1 class of Zintl phases have shown tremendous thermoelectric potential.
Besides, some first-principles simulations have also predicted lucrative
thermoelectric properties, e.g., KSrBi \cite{Wei23} ($ZT\sim2.8$ at 800~K),
NaSrSb ($ZT\sim1.9$), and NaBaSb ($ZT\sim1.0$) at 900~K  \cite{Chandan24}.
This further strengthens our motivation and make SrBaSn a rational
choice for study.

Here, we use state-of-the-art first-principles simulations
integrating density functional theory, harmonic and anharmonic lattice
dynamics, and solutions of Boltzmann transport equation to evaluate
the structural, vibrational, and transport properties of Zintl phases
SrBaX (X = Si, Ge, Sn). As stated earlier, the system of primary interest
is SrBaSn, however, SrBaSi and SrBaGe are also included for a comprehensive
study. These added systems also allow to see the impact of mass contrast and
substitution of heavier atom. Our study not only assess the thermoelectric
potential of proposed systems but also provide key theoretical insights which
could guide future experiments. Some of the highlights of
our study include remarkably low room temperature lattice thermal conductivities
($<1$ W m$^{-1}$ K$^{-1}$), rivaling some of the best materials like
PbTe \cite{Bessas12}, SnSe \cite{Zhao14}, and Bi$_2$Te$_3$ \cite{Wang11}.
We find the origin of such low values in weak bonding in the lattice and
intrinsically suppressed phonon transport, leading to low phonon velocities
and short lifetimes. Further, based on our electronic structure analysis,
we find tremendous thermoelectric performance in SrBaGe, arising from
its suitable electronic structure favoring high Seebeck coefficient
without greatly comprising the electrical conductivity. While SrBaSn
turn out to be most efficient at room temperature, further improvement in
figure of merit is much needed. 

The remainder of the paper is organized in three sections. Section~2
briefly outlines the computational methods used for the study, Sec.~3
discusses results of structural, vibrational, and transport properties,
and finally Sec.~4 summarizes the main findings of the work.

\section{Computational Methods}

We performed first-principles simulations based on density
functional theory (DFT) as implemented in Vienna \textit{ab initio}
simulation package (VASP) \cite{Kresse96, KressePRB} to study the
structural, vibrational, and thermoelectric properties of Zintl
phases SrBaX (X = Si, Ge, Sn). The projector-augmented wave method
was used to treat core and valence electron interactions, combined
with the Perdew-Burke-Ernzerhof (PBE) \cite{Perdew96} generalized
gradient approximation (GGA) for the exchange-correlation functional.
The choice of PBE was based on appropriate balance between computational
accuracy and efficiency. The plane-wave kinetic energy cutoff was set
at 500 eV, after carefully testing the energy and forces convergence
within 10$^{-8}$ eV and 10$^{-7}$ eV/\AA, respectively.

The structural optimization by relaxing all degrees of freedom was performed
using conjugate gradient algorithm. Brillouin zone integration was carried
out using a \textit{k} grid of 11$\times$11$\times$11 for optimization,
while a denser grid of 21$\times$21$\times$ 21 was used for self-consistent
energy calculations, which adequately resolves the electronic states
close to Fermi level. In addition, modified Becke-Johnson (mBJ) potential
\cite{Tran09} was employed for reliable electronic structure, key for
obtaining accurate transport properties. Since the proposed systems
comprise heavy elements like Ba and Sn, relativistic effects were
also included in the electronic structure calculations. To gain
insight into chemical bonding and orbital interactions, we calculated
crystal orbital Hamilton population (COHP) using Lobster code
\cite{Maintz16}.

Harmonic and anharmonic force constants were obtained using finite
displacement method on 2$\times$2$\times$2 supercell (96 atoms),
with a default displacement of atoms by 0.01 \AA. The supercell size
was chosen to reduce finite size effects and considering computational
cost. The resulting phonon dispersions were extracted using Phonopy
code \cite{Togo15} by solving the equation
\begin{equation}
\sum_{\beta\tau'} D^{\alpha\beta}_{\tau \tau'}
(\mathbf{q}) \gamma^{\beta\tau'}_{\mathbf{q}j} =
\omega^2_{\mathbf{q}j}\gamma^{\alpha\tau}_{\mathbf{q}j}.
\end{equation}
where the indices $\tau, \tau'$ represent the atoms, $\alpha, \beta$
are the Cartesian coordinates, ${\mathbf{q}}$ is a wave vector,
$j$ is a band index, $D(\mathbf{q})$ is the dynamical matrix,
$\omega$ is the corresponding phonon frequency, and $\gamma$
is the polarization vector. The lattice thermal conductivity
was evaluated by solving the phonon Boltzmann transport equation
(BTE) within the single-mode relaxation time approximation,
\begin{equation}
\kappa_L = \frac{1}{NV} \sum_\lambda C_\lambda \mathbf {v}_\lambda
\otimes \mathbf {v}_\lambda \tau_\lambda
\label{eqn}
\end{equation}
where $N$ and $V$ are the number of unit-cells in the crystal and
corresponding volume, respectively. The values were converged in terms
of cutoff pair distance (5~\AA) and $q$-point mesh (17$\times$17$\times$17).
To further gain insight into anharmonic phonon scatterings, we calculated
the scattering phase space using ShengBTE code \cite{Sheng14}.

The electrical transport properties were calculated by explicitly
considering acoustic deformation potential (ADP), ionized impurity
(IMP), and polar optical phonon (POP) scattering mechanism as implemented
in Amset code \cite{Ganose21}. The material related parameters such
as elastic constants, dielectric constants, deformation potential,
and phonon frequencies were obtained from first-principles simulations
to calculate carrier scattering rates. Each scattering mechanism's rate was
obtained using Fermi's golden rule \cite{Dirac27}
\begin{equation}
\tilde{\tau}^{-1}_{n \mathbf{k} \rightarrow m \mathbf{k} + q} =
\frac{2 \pi}{\hbar} \arrowvert g_{nm} (\mathbf{k},\mathbf{q})
\arrowvert^2 \delta (\epsilon_{n \mathbf{k}} - \epsilon_{m \mathbf{k} + q})
\end{equation}
where $n$\textbf{k} is the initial wave vector, $m$\textbf{k} +
\textbf{q} is the final waver vector, $\hbar$ is the reduced
Planck’s constant, $\delta$ is the Dirac delta function, $\epsilon$
is the electron energy, and $g$ is the electron-phonon coupling
matrix element. The total carrier scattering rate was obtained
by using the Matthiessen’s rule \cite{Matt64}
\begin{equation}
 \frac{1}{\tau_\mathrm{e}} = \frac{1}{\tau_\mathrm{{ADP}}} + \frac{1}{\tau_\mathrm{{IMP}}} + \frac{1}{\tau_\mathrm{{POP}}}
\end{equation}
The coefficients of electrical transport were further tested for
convergence in terms of interpolation factor, which was found to
be adequate for a default value of 5.

\begin{figure}
\centering\includegraphics[scale=0.45]{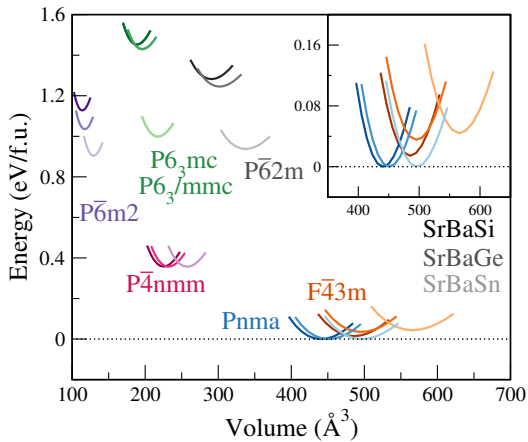}
\caption{Energy versus volume curves for SrBaSi (dark shade), SrBaGe
(medium shade), and SrBaSn (light shade) in different space groups.
The inset shows the enlarged view for orthorhombic and cubic space
groups. Note that the values overlap for P$6_3$/mmc and P$6_3$mc
space groups.}
\label{EvsV}
\end{figure}

\begin{figure}
\centering\includegraphics[scale=0.3]{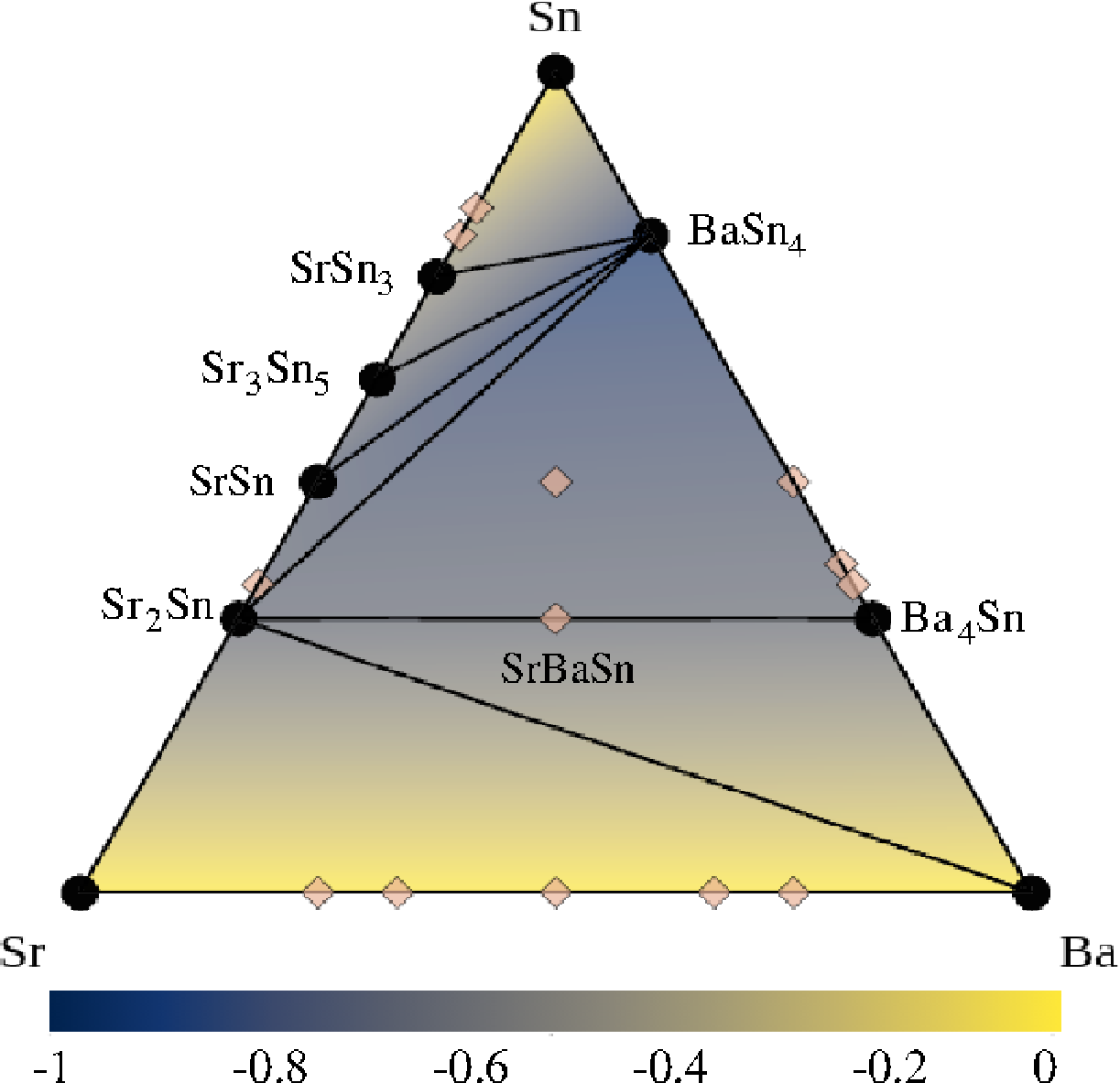}
\caption{Phase diagram for SrBaSn, where colorbar
shows the formation energy (eV/atom), and orange
diamonds represent phases above convex hull.}
\label{Hull}
\end{figure}

\begin{figure}
\centering\includegraphics[scale=0.45]{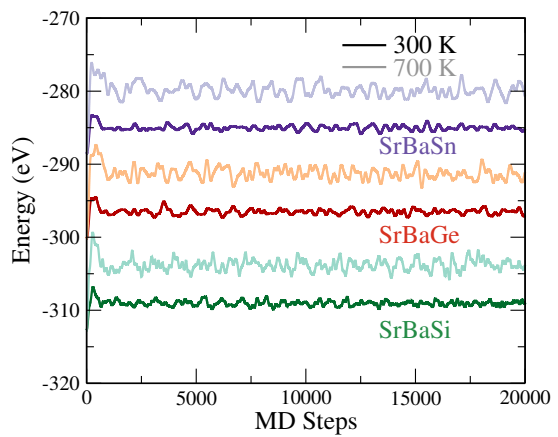}
\caption{Energy evolution for $ab$ $initio$ molecular dynamics
simulations as a function of steps for SrBaX (X = Si, Ge, Sn)
at 300 and 700~K.}
\label{MD}
\end{figure}

\begin{figure*}
\centering\includegraphics[scale=0.43]{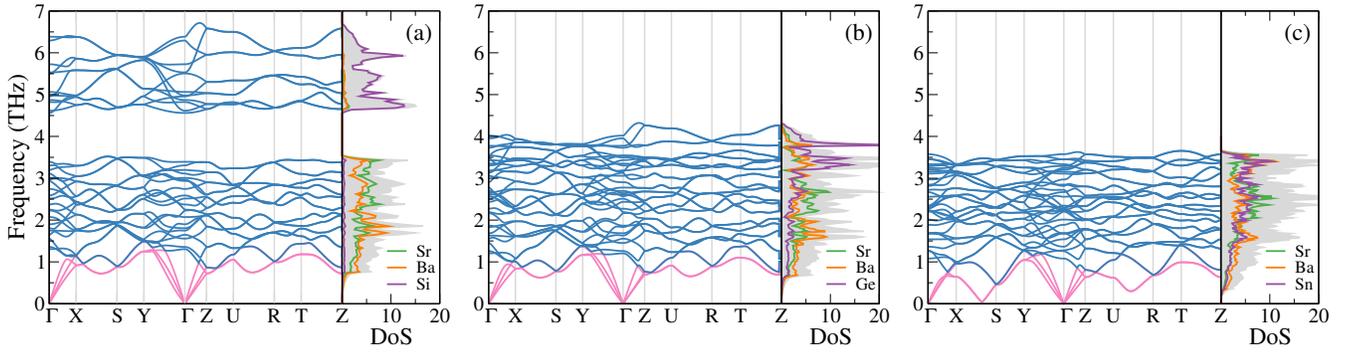}
\caption{Phonon dispersion curves and density of states
(in arb. units) for (a) SrBaSi, (b) SrBaGe, and (c) SrBaSn.}
\label{Phonons}
\end{figure*}

\begin{figure*}
\centering\includegraphics[scale=0.3]{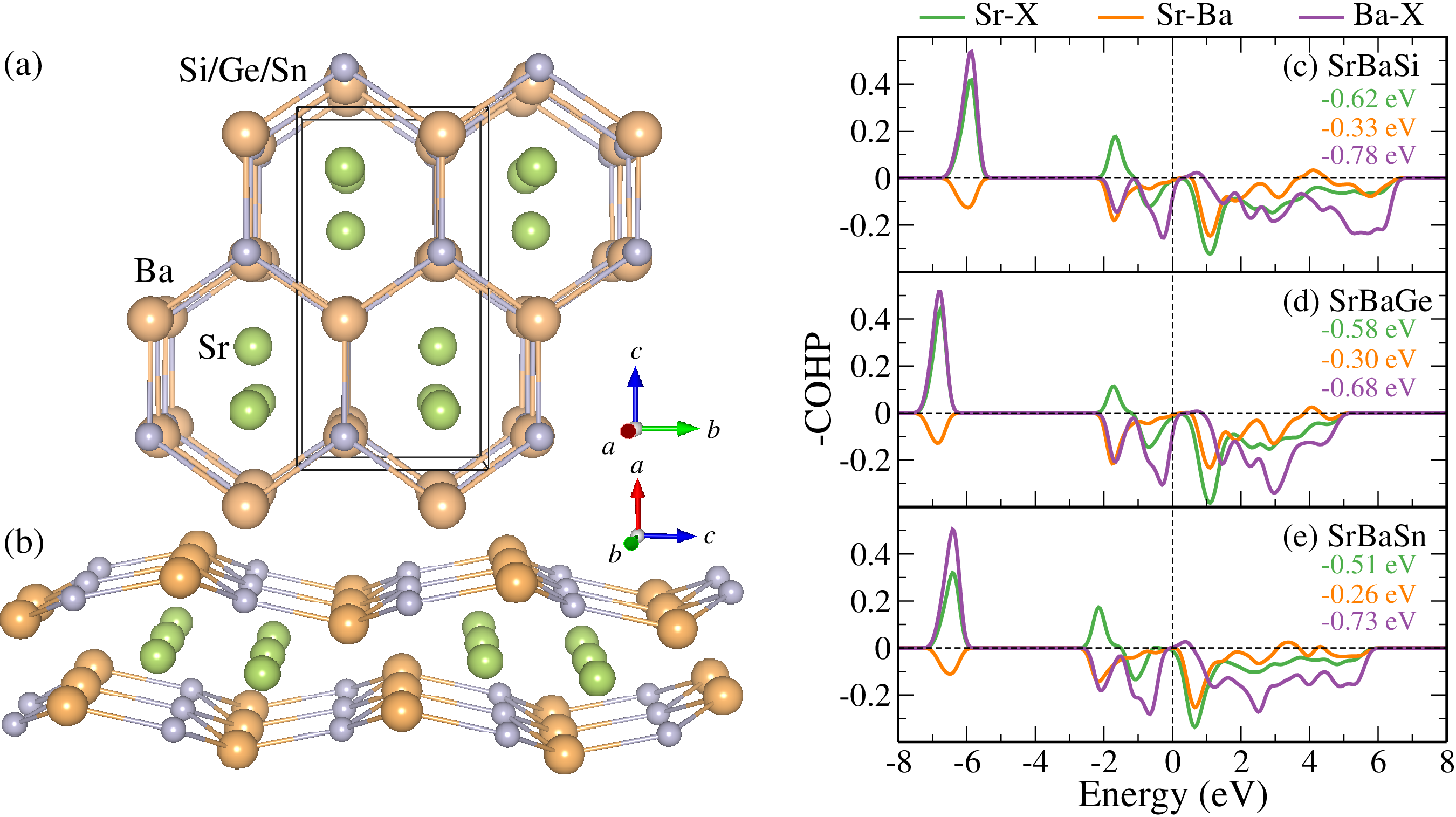}
\caption{(a) Crystal structure of SrBaX (X = Si, Ge, Sn) in
orthorhombic $Pnma$ space group and (b) sideview of the structure
showing puckered hexagonal rings of Ba and X atoms along $bc$-plane,
while Sr atoms are stacked along $a$-axis, (c)-(e) crystal orbital
Hamilton population of SrBaSi, SrBaGe, and SrBaSn, respectively.
The numerical values represent integrated crystal orbital Hamilton
populations for different interaction pairs.}
\label{Crys}
\end{figure*}

\begin{figure}
\centering\includegraphics[scale=0.4]{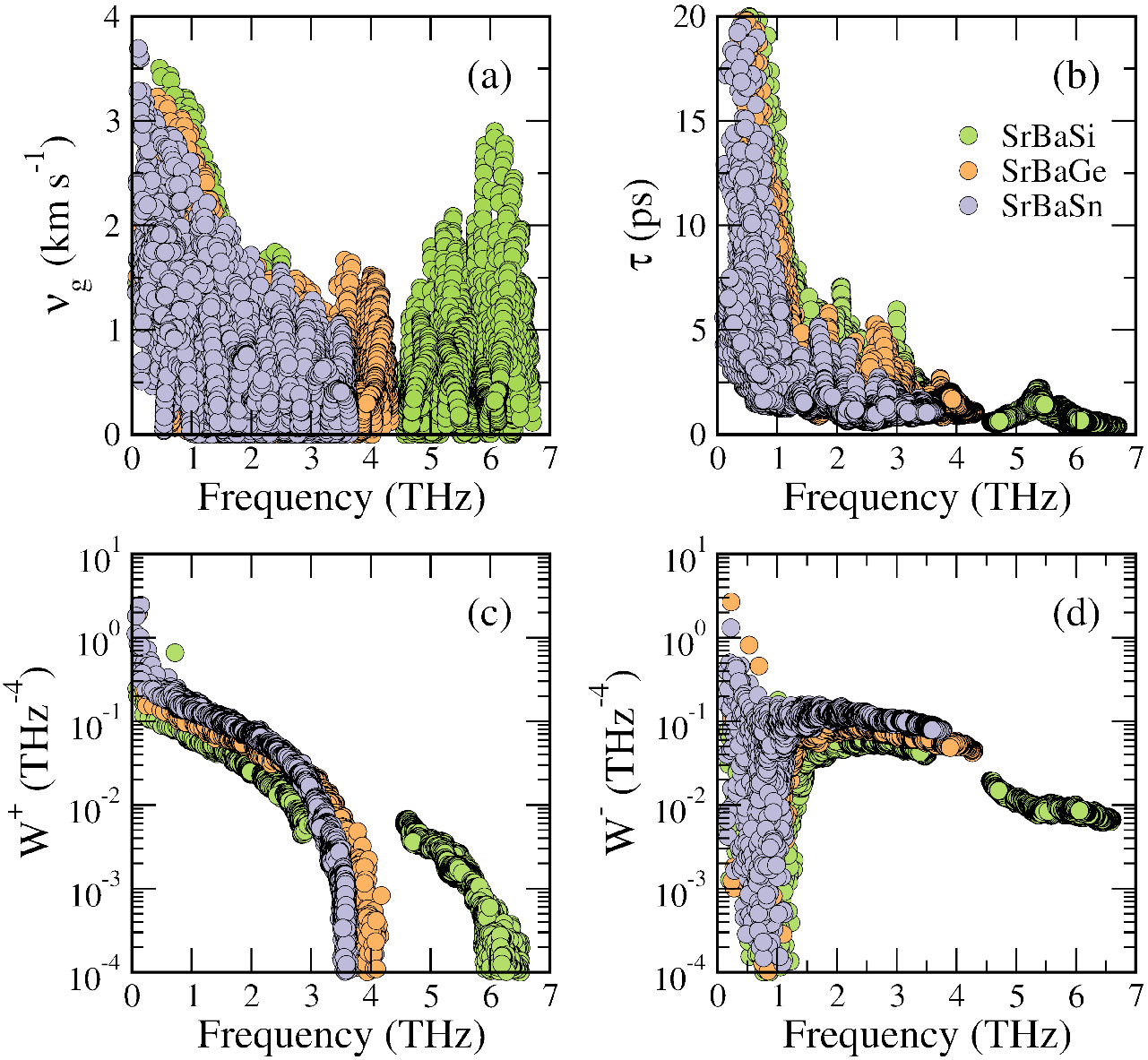}
\caption{(a) Phonon group velocity, (b) lifetime, (c)-(d) absorption
and emission process at 300~K for three-phonon anharmonic scattering
phase space, respectively, as a function of phonon frequency.}
\label{Vg}
\end{figure}

\begin{figure}
\centering\includegraphics[scale=0.42]{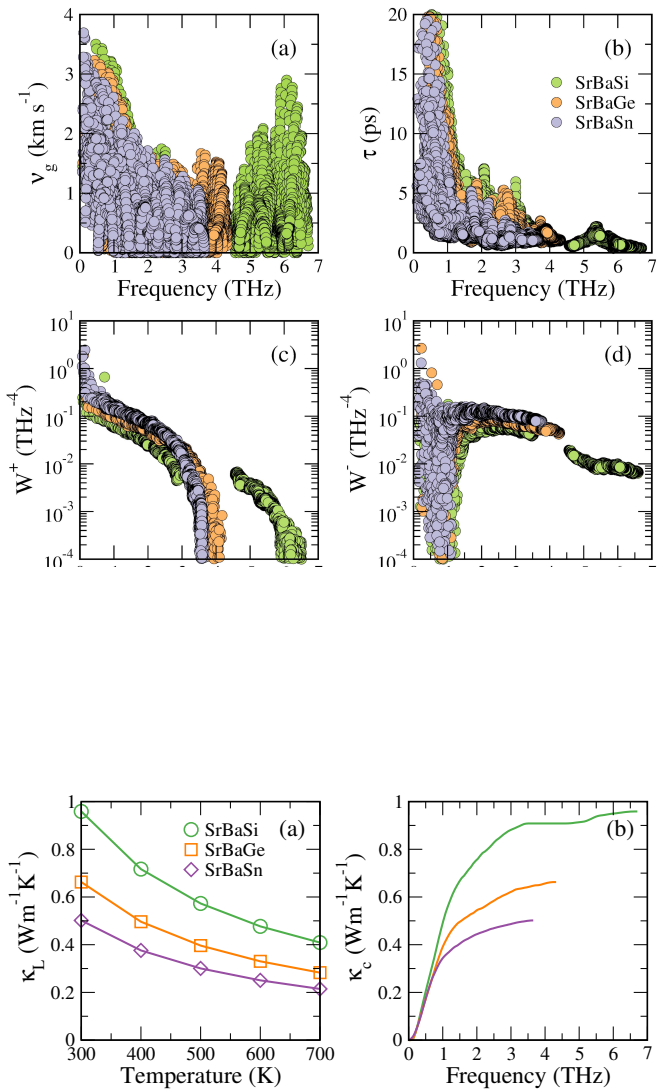}
\caption{(a) Average lattice thermal conductivity as a function of
temperature and (b) cumulative lattice thermal conductivity at 300~K
as a function of phonon frequency.}
\label{Kappa}
\end{figure}

\begin{figure*}
\centering\includegraphics[scale=0.42]{fig8.eps}
\caption{Electronic bands and density of states of (a) SrBaSi,
(b) SrBaGe, and (c) SrBaSn.}
\label{Bands}
\end{figure*}

\begin{figure}
\centering\includegraphics[scale=0.42]{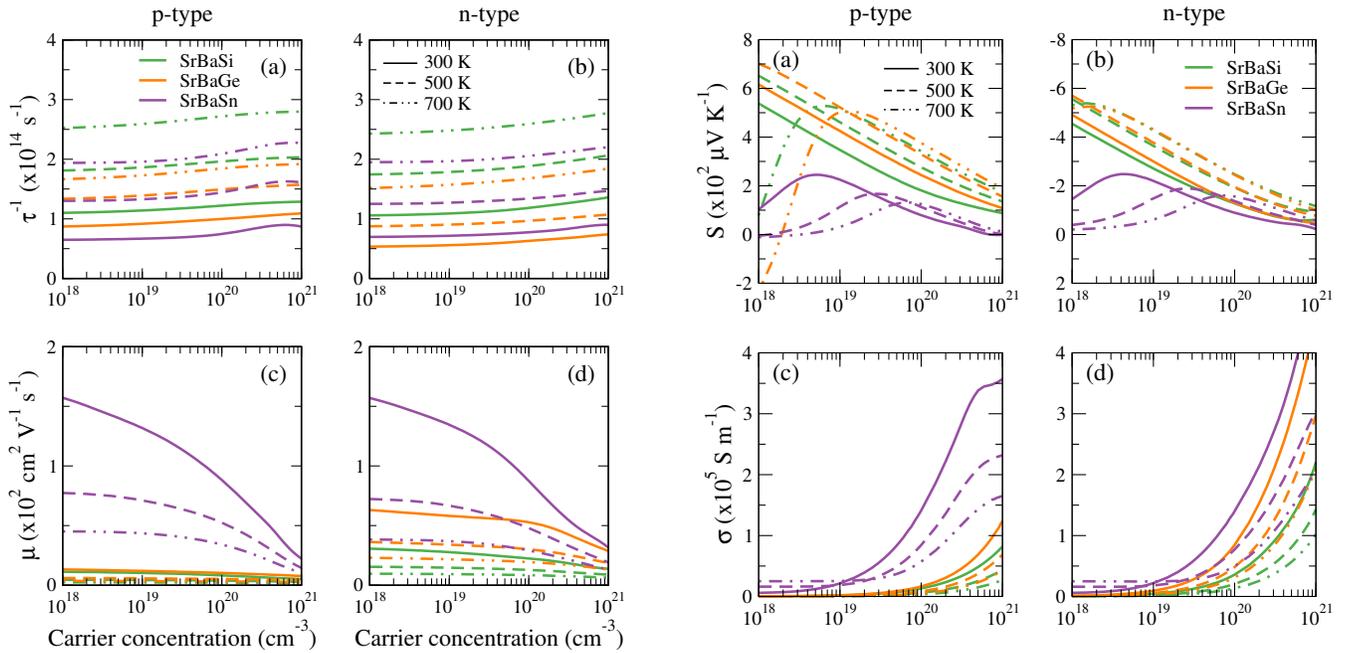}
\caption{(a), (b) Scattering rates and (c), (d) carrier
mobility for $p$-type and $n$-type carrier concentration, respectively,
at different temperatures for SrBaX (X = Si, Ge, Sn).}
\label{Tau}
\end{figure}

\begin{figure}
\centering\includegraphics[scale=0.35]{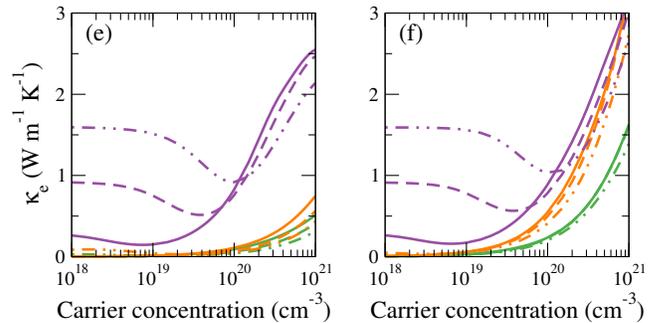}
\caption{(a), (b) Seebeck coefficient, (c), (d) electrical
conductivity, and (e), (f) electronic thermal conductivity,
for $p$-type and $n$-type carrier concentration, respectively,
at different temperatures for SrBaX (X = Si, Ge, Sn).}
\label{PF}
\end{figure}

\begin{figure*}
\centering\includegraphics[scale=0.45]{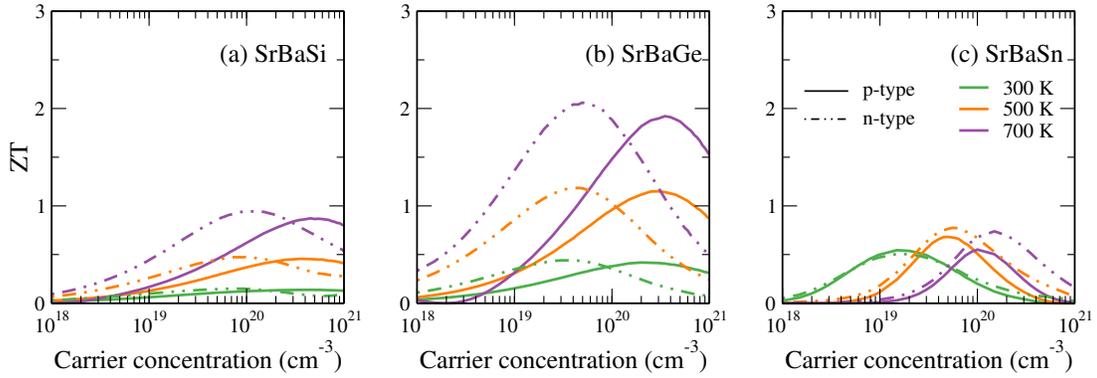}
\caption{Figure of merit as a function of carrier concentration
at different temperatures for $p$-type and $n$-type (a) SrBaSi,
(b) SrBaGe, and (c) SrBaSn.}
\label{ZT}
\end{figure*}

\section{Results and Discussion}

In this section, we discuss the crystal structure, its stability,
phonon dispersions, electronic structure, and transport properties
of SrBaX (X = Si, Ge, Sn) systems.

\subsection{Structural Stability}

SrBaX (X = Si, Ge, Sn), these materials are not experimentally
reported. Therefore, finding the right crystal structure is pivotal.
We stress upon the stability of these materials and address the
question in a hierarchical fashion. Literature reveals that ternary
XYZ type materials containing alkali or alkaline earth metals along
with main group element more often crystallize in cubic, hexagonal,
tetragonal, and orthorhombic symmetries \cite{Bobev23, Zhang12}. The
respective prototype structures are MgAgAs-type, [LiGaGe, ZrNiAl, BaLiSi,
ZrBeSi-type], PbClF-type, and TiNiSi-type. Likewise, we optimized SrBaX
materials in these mentioned space groups. The crystallographic parameters
were adopted from their nearest reported material, e.g., using
SrMgSn \cite{Merlo90} for orthorhombic SrBaSn, among others. The energies of
SrBaX systems in the respective symmetries are shown as a function
of volume in Fig.~\ref{EvsV}. Interestingly, all three systems have
lowest energy in orthorhombic phase. Only the cubic symmetry competes
with orthorhombic to some extent, whereas hexagonal and tetragonal
structures have markedly higher energies. The shape of the energy
well further hints at static stability of the structures.

Ascertaining that SrBaX systems can crystallize in orthorhombic
symmetry, we delve deeper into their stability. We first assess
the thermodynamic favorability in terms of formation energies
($\Delta$E), calculated using the expression
\begin{equation}
 \Delta E = E_{SrBaX} - (E_{Sr} + E_{Ba} + E_X)
\end{equation}
where $E_{SrBaX}$ is the total energy per formula unit, and
$E_{Sr}$, $E_{Ba}$, and $E_X$ are the energies of respective
elements per atom. Our calculated formation energy values are
negative, i.e., $\Delta E$ = -0.90, -1.42, and -1.74 eV/atom for
SrBaSi, SrBaGe, and SrBaSn, respectively, indicating their synthesis
is exothermic and materials are stable with respect to their constituent
elements. More negative value in the case of SrBaSn hints at enhanced
stability under ambient conditions. Though formation energy is a key
ingredient of stability, it merely tells that a material is energetically
favorable. A material can have negative formation energy and could still
decompose into other compositions in the same chemical space.

A true test of thermodynamic stability requires to construct a
convex hull from all known phases within the relevant chemical
space. We constructed convex hull for SrBaX materials using the
OQMD database \cite{Saal13, Kirklin15}, which comprises binary
and ternary phases reported experimentally, along with hypothetical
possibilities based on common structural prototypes. As discernible
from Fig.~\ref{Hull} and Fig.~S1-S2, SrBaSn, and SrBaSi/Ge, respectively,
lie slightly above the convex hull. The respective energy distance from
the hull, $\Delta E_{HD}$, are 0.037, 0.035, and 0.032 eV/atom. By definition,
any phase that sits on the convex hull is thermodynamically stable at 0~K.
Having said that, for materials which do not lie on the convex hull, the
likelihood of synthesis cannot be completely ruled out. In such cases, the
energy distance from the convex hull becomes the defining parameter. For
instance, Ma \textit{et al}. \cite{Ma17} noted that 90 experimentally reported
materials lie close to hull distance within 0.1 eV/atom. It can thus
be surmised that SrBaX may not be strictly stable at 0~K, however,
the small distances from the convex hull do not preclude experimental
synthesis or metastability to say the least. We believe that finite
temperature synthesis can sustain such metastability and these
materials could be experimentally accessible.

To probe the finite temperature stability, we performed $ab$ $initio$
molecular dynamics (AIMD) simulations at 300 and 700~K for SrBaX systems
over 20,000 MD steps, with step size of 0.5 fs. As shown in Fig.~\ref{MD},
at 300~K, the energies across the systems fluctuate narrowly without any drastic
changes. Even at higher temperature of 700~K, when the thermal vibrations
are more pronounced, the oscillations are bounded around the mean position.
Such behavior indicates the structural integrity, robustness, and
unlikeliness of systems to undergo any collapse or irreversible bond
breaking at finite temperature. We further checked the mechanical
stability of the compounds by means of elastic constant values.
For an orthorhombic phase, a stable crystal structure should
follow the Born–Huang elastic stability criteria \cite{Mouhat14},
i.e.,
\begin{subequations}
\begin{equation}
\small \mathrm{C_{11} > 0, C_{11} C_{12} > C_{12}^2}
\end{equation}
\begin{equation}
\small \mathrm{C_{11} C_{22} C_{23} + 2C_{12} C_{13} C_{23}
 - C_{11} C_{23}^2 - C_{22} C_{13}^2 - C_{33} C_{12}^2 > 0}
\end{equation}
\begin{equation}
\small \mathrm{C_{44} > 0, C_{55} > 0, C_{66} > 0}
\end{equation}
\end{subequations}

Our calculated elastic constant values, as listed in Supplemental
Material \cite{Suppl}, obey the stability criteria and highlight
the mechanical stability of these systems. 

In the hierarchy of stability, we lastly checked the dynamical
stability of the systems through phonon dispersions. We observe
from Fig.~\ref{Phonons} that there are no imaginary frequencies
throughout the Brillouin zone for all three systems, indicating
their resistance to any structural distortions. To summarize
thus far, based on energy profile, formation energy, convex hull
analysis, AIMD simulations,  and phonon dispersions, we have
demonstrated the stability of SrBaX (X = Si, Ge, Sn) systems
in orthorhombic phase. Now we proceed to look at the crystal
structure arrangement and the bonding features in orthorhombic
symmetry.

\subsection{Crystal Structure and Bonding}

The crystal structure of SrBaX (X = Si, Ge, Sn) in orthorhombic
symmetry emulates the well-known TiNiSi structural prototype in
$Pnma$ space group. The TiNiSi-type structure has been discussed
in lengths in literature \cite{Landrum98}. Here, we briefly recapitulate
the essential structural aspects central to the theme of our work.
The crystal structure has puckered six-membered rings of Ba-X
stacked along $yz$-plane, whereas Sr-atoms are centered along
$x$-axis between two such hexagonal rings, as shown in Fig.~\ref{Crys}.
The nearest neighbor distance for X-X atoms is of the order of
$\sim$5~\AA, suggesting no bonding interactions among them and
ruling out any possibility of anionic framework for [X-X]$^{n-}$.
This is an indication of [Ba-X]$^{n-}$ type network, whereas the Sr
atoms are weakly linked to the rest framework. To gain a better
understanding, we performed the -COHP analysis for predicting
bonding features in a crystal lattice.

The -COHP interactions for Sr-Ba, Sr-X, and Ba-X pairs are shown
in Fig.~\ref{Crys}(c)-(e). In all three systems, bonding interactions
are prevalent well below the Fermi level. The dominant interactions are
particularly for Ba-X, followed by Sr-X, as can be noticed by large positive
-COHP peaks in the region of -5 to -8 eV for different systems. The integrated
COHP (ICOHP) values provide a quantitative measure of bonding strength and
further substantiates the stronger bonding interactions in Ba-X (-0.78 eV,
-0.68 eV, and -0.73 eV, respectively). This suggests the [Ba-X]$^{n-}$ type
framework in the lattice, whereas Sr-atoms can act as electropositive
entity donating charge to the lattice, consistent with the Zintl
concept. Furthermore, the decreasing ICOHP values demonstrate gradual
reduction in bonding strength on going from SrBaSi to SrBaSn,
highlighting the relative phonon softening and suppressed
thermal transport.

Notably, there are prominent antibonding states for Ba-X
and Sr-X interactions just below the Fermi level in the
valence band maxima, indicating weak bonding in the lattice.
Such antibonding states result in softer phonon modes \cite{Wei23}
and reduced phonon group velocities, as discussed in the
following section.


\subsection{Phonons and Lattice Thermal Conductivity}

In addition to dynamic stability, as noticed from
Fig.~\ref{Phonons}, now we revisit phonon dispersions
from the viewpoint of thermal transport. In total, there
are 36 phonon modes per wave vector, originating from 12 atoms
in the primitive unit cell. Generally, the phonon modes are
organized hierarchically in order of atomic masses of constituent
elements. In a broader aspect, the lighter atoms (Si, Ge)
contribute to the high-frequency optical modes, whereas
heavier atoms (Sr, Ba) participate dominantly in the
acoustic modes for SrBaSi/Ge. Since the size difference of
constituent elements in SrBaSn is not as substantial as rest
materials, it displays a mixed behavior. For instance, the Sn
atoms have a greater say in high-frequency phonon modes, whereas
the acoustic modes are dominated by Ba atoms, with significant
contributions of Sr and Sn atoms as well.


There is a noticeable bunching of phonon modes as one go
from SrBaSi (6.5~THz) to SrBaGe (4.5~THz), and to SrBaSn
(3.5~THz), pointing at smaller slopes of phonon modes.
Since the slope of the phonon modes correspond to phonon
group velocities, i.e., $\nu_g = d \omega/dq$, all materials
are expected to have low phonon velocities, as shown in
Fig.~\ref{Vg}(a). The respective average sound velocities
are significantly low, i.e., 2.34, 2.20, and 2.02 km~s$^{-1}$,
which puts them in-line with renowned materials such as
SnSe \cite{Zhao14}, PbTe \cite{Bessas12}, and Bi$_2$Te$_3$
\cite{Wang11}. The $\nu_g$ shows a decreasing
trend with phonon frequencies on going from Si to Sn,
depicting the effect of increasing atomic mass and
weak bonding in the lattice. In SrBaSi, acoustic modes
reaches $\sim$3 km~s$^{-1}$, the mid-frequency phonon modes
span close to $\sim$1.5 km~s$^{-1}$, and the high
frequency modes above the phonon gap also have
reasonable contribution, consistent with the dispersive
nature of phonons and hints at stronger interactions
among atoms. SrBaGe has slightly lower $\nu_g$ values
and the dispersions are confined to 4.3~THz, reflecting
the softening of phonon modes. In SrBaSn, the majority
of acoustic modes lie below $\sim$2~km~s$^{-1}$, whereas
the most optical modes are clustered below 1~km~s$^{-1}$,
correlating with the flat phonon modes. The reduced $\nu_g$
in SrBaSn is testament to intrinsically weak phonon transport
and pronounced acoustic-optical coupling, also indicating short
phonon lifetimes.

The frequency dependent phonon lifetimes ($\tau$) of SrBaSi,
SrBaGe, and SrBaSn are presented in Fig.~\ref{Vg}(b). The lifetimes
decrease rapidly with frequencies due to enhanced anharmonic
phonon-phonon scatterings. The majority of phonon modes in the
region below 1.5~THz have lifetimes within 10~ps, reminiscent
of distinguished material SnSe \cite{Guo15}. In particular, SrBaSi depicts
relatively longer lifetimes across the frequency range, consistent
with its more dispersive phonon modes in comparison to rest materials.
Notably, SrBaSn exhibits the shortest lifetimes with most modes lying
well below 4~ps across the spectrum, in agreement with its relatively
flat phonon modes, low group velocities, and strong acoustic-optical
mixing. The lifetimes of SrBaGe show an intermediate behavior and lie
approximately in between those of SrBaSi and SrBaSn. Importantly, such
low lifetimes suggest pronounced anharmonicity in the crystal lattice.

To delve deeper into thermal transport, we next study the
three-phonon anharmonic scattering phase spaces for SrBaX
systems for absorption (W$^+$) and emission (W$^-$) processes
as shown in Fig.~\ref{Vg}(c)-(d). For low frequency phonon modes
($<$ 1.5~THz), W$^+$ is higher, suggesting dominance of
absorption processes for acoustic modes. Across the materials,
W$^+$ decreases rapidly with frequency, showing the typical
inverse relation between scattering and frequency. SrBaSn
exhibits the largest W$^+$ values for most of the region,
substantiating its short phonon lifetimes. On the other hand,
SrBaSi shows lowest values, indicating relatively weak
phonon-phonon interactions and accordingly higher lifetimes.
Yet again, SrBaGe displays intermediate values. The shorter
W$^+$ for optical modes, especially high frequency ones,
corroborates that heat transport is predominantly governed
by the acoustic modes. Further, W$^-$ gradually increases
with frequency and peaks around 1.5~THz, indicating
acoustic-optical mixing. Thereafter, W$^-$ consistently
decreases with increasing frequency. Once again SrBaSn show
larger W$^-$, SrBaSi lowest, and SrBaGe lies in between.
Overall, the analysis of anharmonic scattering phase spaces
points at weaker thermal transport in SrBaSn in comparison
to other counterparts.

The collective behavior of phonon group velocities, lifetimes,
and scattering phase spaces implies severely hindered thermal
transport in SrBaSn in comparison to SrBaSi, whereas SrBaGe
shows moderate behavior. Likewise, as discernible from
Fig.~\ref{Kappa}, we obtained the trend of average $\kappa_L$
in the order of SrBaSi $>$ SrBaGe $>$ SrBaSn. As a matter of
fact, the $\kappa_L$ values are nearly half for SrBaSn in
comparison to SrBaSi. Further, the $\kappa_L$ decreases with
increasing temperature on account of enhanced phonon-phonon
scatterings at higher temperatures. It is interesting to note
that the $\kappa_L$ values range 0.95-40, 0.66-0.28, and 0.50-0.21
W m$^{-1}$ K$^{-1}$ in the temperature region of 300-700~K
for SrBaSi, SrBaGe, and SrBaSn, respectively. These numbers
correlate fairly well with well-known thermoelectric materials
like SnSe ($\kappa_L \sim 0.47$ W m$^{-1}$ K$^{-1}$) \cite{Zhao14},
PbTe \cite{Bessas12} ($\kappa_L \sim 2.0$ W m$^{-1}$ K$^{-1}$), and
Bi$_2$Te$_3$ \cite{Wang11} ($\kappa_L \sim 1.6$ W m$^{-1}$ K$^{-1}$),
and also with some other 1-1-1 type Zintl phases such as BaCuSb
\cite{Zheng22} and NaCdSb \cite{Guo23}, suggesting the thermoelectric
potential of studied systems, especially SrBaSn.

Such phenomenally low values intrigued us to further gain
insights into atomic contributions toward $\kappa_L$. Hence,
we calculated cumulative $\kappa_L$ as a function of phonon
frequency. At 300~K, the low frequency phonon modes up to
2.0~THz contribute extensively to the $\kappa_L$, i.e.,
~88\% in SrBaSn to 80\% in SrBaSi. In conjunction with
Fig.~\ref{Phonons}, we observe that these modes are
predominantly contributed by Sr and Ba atoms in
SrBaSi/Ge, whereas Ba has major contribution in
the case of SrBaSn. Thus, any attempt to lower the
$\kappa_L$ should target doping at alkaline earth metal
(Sr, Ba) sites. The favorable dopants could be lighter
atoms like Y and Ce, acting as scattering centers which
could disrupt the phonon propagation. For instance, Tihtih
\textit{et al}. \cite{Tihtih23} demonstrated the reduction of $\kappa_L$ in
Sr and Y-doped BaTiO$_3$. Further, the Sr-rich or Ba-deficient
compositions can also help in lowering the $\kappa_L$, as observed
in another study. Sr-doped BaTiO$_3$ showed lattice disorder
facilitating the scattering of phonons \cite{Khalil24}.

Altogether, these materials demonstrate hindered thermal transport
and invites further exploration of electrical transport properties
from the viewpoint of thermoelectric performance. The electrical
transport properties are derived exploiting electronic structure
and scattering rates, as discussed in the following section.

\subsection{Electronic structure and Transport Properties}

The electronic band structures and density of states of SrBaX
are presented in Fig.~\ref{Bands}. All the materials are
semiconducting in nature, beneficial from the perspective
of thermoelectrics. While only the lighter composition SrBaSi
displays direct bandgap of 0.69 eV along $\Gamma$-point, SrBaGe
shows an indirect bandgap of 0.66 eV, whereas the value is
drastically reduced to 0.11 eV in the case of heavier SrBaSn.
Such low bandgap may promote bipolar conduction in SrBaSn.
Though all the materials show similar overall
topology, but there are slight variations in the position and
curvature of the bands near the Fermi level, which have direct
relevance to the transport properties. All systems have more
dispersive bands in the conduction band region in comparison
to valence band region. Thus, $p$-type systems are likely to
exhibit higher Seebeck coefficients, whereas higher electrical
conductivities for $n$-type systems. SrBaSi and SrBaGe have
relatively more flat bands in the valence band region in
comparison to SrBaSn, indicating larger Seebeck coefficients
for $p$-type systems. Meanwhile, the conduction band minima
of SrBaSn has relatively more dispersive nature of bands in
comparison to others. Thus, in conjunction with reduced bandgap,
SrBaSn is likely to exhibit enhanced electrical conductivities
in comparison to other counterparts.

It is interesting to note that in addition to valence band
maximum (VBM), there is a second valley along $\Gamma$-Z
direction within 0.05 eV energy window in all the cases.
Such multiple valleys facilitate the mobility of charge
carriers and improve power factor \cite{Tang15}. To shed more light on
the electronic states near the Fermi level, we further
analyzed the projected density of states (DoS). The states
close to Fermi level in the valence band maxima are majorly
contributed by X atoms (X = Si, Ge, Sn), consistent with their
greater electronegativity, whereas Sr and Ba atoms dominate
in the conduction band minima. The DoS features further
compliment the analysis of electronic bands. SrBaSi and SrBaGe
exhibit sharper, more localized DoS at the valence band edges,
whereas SrBaSn displays a broader distribution. These observations
in SrBaSi and SrBaGe suggest larger effective mass and more
localization of carriers close to valence band extrema, while
SrBaSn promotes lighter carriers with enhanced velocities.

Based on the electronic structure, our calculated net
scattering rates and mobility of charge carriers are
shown in Fig.~\ref{Tau}. The various components of
scattering rates, viz., ADP, IMP, and POP, as shown
in Fig.~S8, demonstrate the dominance of POP scattering
for SrBaSi and SrBaGe at considered temperatures (300, 500,
and 700~K). While SrBaSn show similar behavior at moderate
temperatures, ADP is the dominating mechanism at 700~K.
This shows that, except for the latter case, charge
carriers interact more with polar optical phonons.
Nonetheless, the net scattering rates increase
steadily with carrier concentrations and rapidly
with temperature for both $p$- and $n$-type dopings.
Overall, the net scattering rates range 0.8-2.7 fs$^{-1}$
and 0.7-2.7 fs$^{-1}$ for holes and electrons, respectively.

Now, we move to the other aspect of Fig.~\ref{Tau},
i.e., mobility of charge carriers, which play an
important role in understanding the transport behavior.
Mobility of carriers is a cumulative effect of scattering
rates, band dispersions, and effective mass of charge carriers.
For SrBaSi and SrBaGe, electrons show higher mobilities in
comparison to holes, on account of more dispersive bands
in conduction band minima. On the other hand, SrBaSn has
similar mobilities for both holes and electrons, consistent
with dispersive bands near either band edges. The notable
observation is the standout anomalously high mobility
of SrBaSn, originating collectively from its small bandgap
(0.11 eV) and highly dispersive bands close to band extrema.
Further, the decreasing mobility trend with carrier
concentration and temperature is consistent with frequent
scatterings and increased thermal motion of charge carriers.
Following this analysis, we next evaluate the resulting
electrical transport properties.

As discernible from Fig.~\ref{PF}, from a broader
perspective, SrBaGe exhibits largest Seebeck coefficients
at 300 and 500~K, SrBaSi show larger values at 700~K
for small carrier concentrations, whereas SrBaSn
has lowest numbers. This can be attributed to flat
bands and larger density of states close to band edges
in SrBaSi and SrBaGe with respect to SrBaSn,
as illustrated in Fig.~\ref{Bands}. At 300 and 500~K,
Seebeck coefficient for both $p$- and $n$-type
SrBaSi and SrBaGe decreases monotonically with
increasing carrier concentrations as the Fermi
level moves deeper into the bands. The trend is
also consistent with Mott equation for doped
semiconductors \cite{Heremans08, Cutler69}. However,
at 700~K, a pronounced maximum at low to moderate carrier
concentrations is observed for holes, whereas a tiny glimpse
of the same can be seen for electrons. Such behavior
at high temperatures is often associated with bipolar
conduction, especially in materials with narrow band gap,
for instance, SrBaSn (E$_g$ $\sim$ 0.11 eV) in the present
study. Unlike SrBaSi and SrBaGe, SrBaSn demonstrate bipolar
effects throughout the considered temperatures, for both holes
and electrons.  

In contrast, electrical conductivity benefits from such phenomenon
because both electrons and holes can jointly contribute,
as can be noticed from Fig~\ref{PF}. SrBaSn has phenomenally
higher $\sigma$ in comparison to SrBaSi and SrBaGe. The
trend is in agreement with the dispersive bands
and significantly higher mobilities of charge carriers.
In general, the electrons have higher conductivities
in SrBaSi and SrBaGe, whereas SrBaSn exhibit comparative
values for either type of charge carriers. The behavior
correlates with the mobilities of charge carriers, as
discussed in the context of Fig.~\ref{Tau}. Overall,
the $\sigma$ increases with carrier concentration on
account of enhanced carriers, and values decrease
consistently with temperature due to increased collisions.

However, it should be acknowledged that the downside
of bipolar conduction is often elevated electronic thermal
conductivities which may harm the overall performance of the
material, as observed in the present case. The $\kappa_e$ of
SrBaSi and SrBaGe closely follows the trend of $\sigma$. However,
$\kappa_e$ in SrBaSn shows a non-monotonic dependence on carrier
concentration on account of bipolar effect. The values are nearly
constant at the low doping level, then decreasing over an intermediate
dopings, and finally increasing at higher dopings. Notably, SrBaSn has
significantly higher $\kappa_e$ in comparison to other counterparts,
which may compensate its phenomenally high $\sigma$ values. The
evolution of $S$, $\sigma$, and $\kappa_e$, with temperature and
carrier concentration governs the overall thermoelectric performance,
as discussed next. 

\subsection{Thermoelectric Performance}

Figure~\ref{ZT} shows the figure of merit values with respect
to carrier concentrations at different temperatures for SrBaSi,
SrBaGe, and SrBaSn. The conflicting trends of $S$, $\sigma$, and
$\kappa_e$ render a peak in $ZT$ values. In general, $ZT$ increases
with temperature in SrBaSi and SrBaGe, whereas the values decrease
in SrBaSn beyond 500~K. SrBaGe emerges as standout material exhibiting
the highest thermoelectric performance at moderate and high temperature.
It attains $ZT\sim$ 1.1 at 500~K, and $ZT\sim$ 1.9 and 2.0 at 700~K,
for $p$- and $n$-type dopings, respectively, surpassing both SrBaSi
and SrBaSn. It is interesting to note that our calculated values
of SrBaGe are competitive with other state-of-the-art materials
such as SnSe ($ZT\sim$ 2.6 at 913~K), PbTe-SrTe ($ZT\sim$ 2.5 at
923~K), CoSb$_3$ ($ZT\sim$ 1.7 at 823 K), and other Zintl phases
of the same family, e.g., NaCdSb ($ZT\sim$ 1.3 at 673~K), and
BaCuSb ($ZT\sim$ 0.4 at 1010 K). Nevertheless, the superior performance
of SrBaGe can be attributed to delicately balanced electronic structure,
resulting in large $S$ without severely comprising $\sigma$. In contrast,
SrBaSi shows moderate enhancement in $ZT$ with temperature and reaches values
close to unity at 700~K. On the other hand, SrBaSn show less
improvement in $ZT$ on going from 300 to 500~K, and further undergo
decline in $ZT$ at 700~K. The collapse in $ZT$ at high temperature can
be attributed to its small band gap, leading to substantial bipolar effects.
Thus, despite having large $\sigma$ values, SrBaSn suffers the simultaneous
rise of $\kappa_e$. However, SrBaSn show highest values at 300~K
among other materials, i.e., $ZT$ $\sim$ 0.5.

SrBaSn could be a good thermoelectric prospect at room temperature
(300~K), provided its figure of merit could be improved to some extent.
We believe that alloying SrBaSn with SrBaGe could help in reducing the
bipolar conductivity through bandgap engineering. A modest improvement
in bandgap may suppress the larger values of $\kappa_e$. Similarly, the
performance of SrBaSi can be improved through alloying with SrBaGe,
introducing strain in the system \cite{Guo18}, or applying hydrostatic
pressure \cite{Javed17}. Such techniques can lead to band degeneracy, which
has been fruitful in improving power factor. In addition, SrBaSi has scope
for reducing $\kappa_L$ through nanostructuring. Besides, an interesting observation
is the relatable $ZT$ values for both $p$- and $n$-type systems, suitable
from the viewpoint of device fabrication. In a thermoelectric module,
it is desirable to have both legs of similar material, or ideally of
the same material, which is rare. This shows the importance of the
SrBaX series in the context of practical applications.

The optimal carrier concentrations for desirable figure of merit
values for SrBaGe are of the order of 10$^{19}$-10$^{20}$ carriers
cm$^{-3}$ for $n$-type dopings, whereas slightly higher carrier
concentrations are required for $p$-type dopings. However, Zintl
phases have an intrinsic disposition towards $p$-type dopings
\cite{Brown06, Wang07}. It is worth mentioning that Toberer
\textit{et al}. \cite{Toberer09} reported $p$-type
LiZnSb with carrier concentrations of the order of 10$^{20}$ carriers
cm$^{-3}$. The real challenge could be achieving $n$-type carriers,
however, extrinsic dopings could be the pragmatic way ahead. The
plausible ways could be introducing aliovalent substitution at
cation or X-site. Realistic approach may include substitution
on the Sr/Ba-site with trivalent donors such as La$^{3+}$, Ce$^{3+}$,
and Y$^{3+}$, or substitution on the group-14 X-site by group-15
dopants (e.g., Sb/Bi on Ge). Similar approaches can be exercised
for SrBaSi and SrBaSn. We hope that despite the theoretical nature
of present study, our proposed results and recommended guidance
would invite further experimental investigations in order to explore
the thermoelectric potential of the SrBaX series. 

\section{Conclusion}

In this paper, with an aim to design new thermoelectric materials,
we have investigated the structural, vibrational, and transport
properties of SrBaX (X = Si, Ge, Sn) using state-of-the-art
computational methods which combining density functional theory,
harmonic and anharmonic lattice dynamics, and solutions of
Boltzmann transport equation. Since the chosen materials are
not experimentally reported, we rigorously tested their stability
through the energy profile, formation energy, convex hull analysis,
$ab$ $initio$ molecular dynamics simulations, and phonon dispersions.
All materials were stabilized in orthorhombic symmetry and show
semiconducting behavior (E$_g\sim$ 0.11-0.69 eV). The -COHP analysis
revealed weak bonding in the lattice, conducive to softer phonon
modes and reduced phonon group velocities. Our analysis of phonons
resulted in weak phonon transport, leading to low phonon velocities,
short lifetimes, and enhanced scattering phase space, which are
detrimental to thermal conductivity. Likewise, we obtained remarkably
low lattice thermal conductivities ($<$1~W~m$^{-1}$~K$^{-1}$),
correlating with well-known thermoelectric materials like SnSe.
Based on such low values and combining with electrical transport
properties, we obtained tremendous thermoelectric potential in
SrBaGe ($ZT\sim$ 2.0 at 700~K). While SrBaSn demonstrated
anomalously high electrical conductivities, the bipolar conduction
severely affected its overall performance. SrBaSi showed $ZT$ values
close to unity at 700~K. Nonetheless, the favorable $ZT$ values on
either type of dopings, make these materials suitable for device
fabrication. We believe that further experimental investigations
based on our study would unearth new potential thermoelectric
prospects and could serve as motivation for more such studies. 

\begin{acknowledgements}
 
V.G. and M.Z. are thankful to BUKRP00034052 and SERB-DST (File No. PDF/2022/002559),
respectively, for the financial assistance. B.K.M. acknowledges the
funding support from the SERB, DST (CRG/2022/000178). The calculations
were performed using the PARAM Rudra, a national supercomputing facility
at the Inter-University Accelerator Centre (IUAC) New Delhi.

\end{acknowledgements}

\bibliography{SrBaX}

\end{document}